\documentclass[aps,prl,superscriptaddress,10pt,twocolumn,amsmath]{revtex4}

\usepackage{graphicx}

\begin{document}

\title{Biogenic crust dynamics on sand dunes}

\author{Shai Kinast}
\affiliation{Department of Solar Energy and Environmental Physics, Blaustein Institutes for Desert Research, Ben-Gurion University of the Negev, Sede Boqer Campus, 84990, Israel}
\author{Ehud Meron}
\affiliation{Department of Solar Energy and Environmental Physics, Blaustein Institutes for Desert Research, Ben-Gurion University of the Negev, Sede Boqer Campus, 84990, Israel}
\affiliation{Department of Physics, Ben-Gurion University, Beer Sheva, 84105, Israel}
\author{Hezi Yizhaq}
\author{Yosef Ashkenazy}
\affiliation{Department of Solar Energy and Environmental Physics, Blaustein Institutes for Desert Research, Ben-Gurion University of the Negev, Sede Boqer Campus, 84990, Israel}

\date{\today}


\begin{abstract}
Sand dunes are often covered by vegetation and biogenic
crusts. Despite their significant role in dune stabilization, biogenic
crusts have rarely been considered in studies of dune dynamics. Using
a simple model, we study the existence and stability ranges of
different dune-cover states along gradients of rainfall and wind
power. Two ranges of alternative stable states are identified: fixed
crusted dunes and fixed vegetated dunes at low wind power, and fixed
vegetated dunes and active dunes at high wind power. These results
suggest a cross-over between two different forms of desertification.
\end{abstract}

\maketitle

Sand dunes have been the subject of active research for many years,
largely because of their fascinating shapes and dynamics
\cite{bagnold,fryberger,tsoar_book,duran_dunes_review}. Current
studies have increasingly addressed the question of sand-dune stability in
relation to climate change and anthropogenic disturbances
\cite{thomas_2005_remobilization,ashkenazy_2012_climatic_change,provoost_2011_dunes_europe}.
Sand dunes are considered ``stable'' when
they are fixed in place or are stationary~\footnote{We distinguish here
between \emph{stable dunes} and \emph{stable dune states}. The
former term refers to fixed or stationary dunes, whereas the latter
refers to asymptotic stability in the sense of dynamical-system
theory.}. Their stability is strongly affected by the degree of
vegetation coverage. High coverage reduces the wind power at the dune
surface and thereby acts to immobilize the dunes. The re-mobilization
of fixed dunes, either by vegetation mortality or clear-cutting, often
has detrimental effects on the unique ecosystems that develop in
stable dunes \cite{brown_species_1973,tsoar_2005}, leading to
alternative ecosystems associated with active sand \cite{ziv_2006_gerbils}. Active dunes may also pose a threat to human settlement as they can block roads and cover residential areas and agricultural fields \cite{dong_2004_roads,khalaf_1993_kuwait}.

Sand dunes are also stabilized by biogenic soil crusts. These crusts comprise a variety of organisms, including cyanobacteria, lichens and mosses, which live at the surface of desert soils \cite{belnap_book}. Biogenic crusts enhance the aggregation of sand grains, prevent saltation, and reduce wind erosion. Since most sandy soils are located in drylands where the vegetation is patchy and generally sparse \cite{tsoar_book}, the role of biogenic crusts in stabilizing dunes is important and often crucial \cite{veste_2001_crust_role}. Despite their significance and vast presence in the Kalahari, Australian and Central Asia deserts \cite{Thomas_2007_kalahari,Hesse_2006_australia,orlovsky_2004}, soil crusts have rarely been considered in studies of dune dynamics \cite{yizhaq_prl,herrmann_veg_dunes}.

\begin{figure}[h]
  \centering
  \includegraphics[width=0.45\textwidth,height=8.0cm]{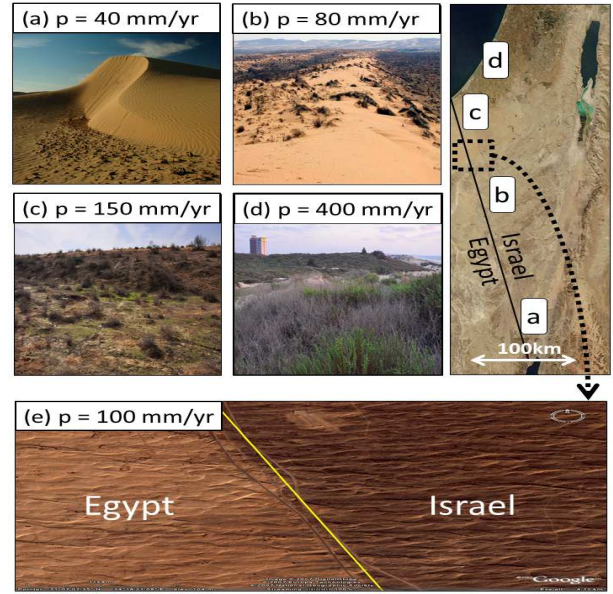}
  \caption{Different types of sand dune cover along a rainfall gradient in Israel (\textit{p} denoting the mean annual precipitation): (a) bare and active dune in hyper-arid region (Kasuy sands); (b) partially crusted and vegetated semi-active dune in arid     region (Nizzana, 50\% crust cover and 10\% vegetation cover~\cite{raz_thesis}); (c) almost fully crusted and vegetated fixed dune in arid region (Halutza, 80\% crust cover and 15\% vegetation cover~\cite{yair_gradient_2008}); (d) vegetated fixed dune in semi-arid region (Nitzanim); (e) the Israel-Egypt border region with crusted fixed dunes on the Israeli side alongside bare and mobile dunes on the Egyptian side.}
  \label{fig:dunes_pgradient}
\end{figure}

Depending on wind power and precipitation level different dune-cover
states are observed. Figure \ref{fig:dunes_pgradient} shows several
typical states from regions of relatively weak winds.
At very low precipitation levels (Fig. \ref{fig:dunes_pgradient}a),
dunes are active due to low crust and vegetation coverage. As the
precipitation increases, the dunes are gradually stabilized. At low
precipitation levels, the dominant stabilizing agent is the soil crust
(Fig. \ref{fig:dunes_pgradient}b,c), while at high levels, the
stabilizing agent is predominantly vegetation
(Fig. \ref{fig:dunes_pgradient}d). Although vegetation and biogenic
crusts have similar effects in stabilizing dunes, they lead to
ecosystems that differ vastly in their bioproductivity.

Motivated by these observations, we ask whether the transition from
crusted to vegetated dunes along the rainfall gradient is gradual or
abrupt, and how it is affected by the wind power. Studying these
questions is significant for understanding desertification processes,
i.e. processes involving the irreversible loss of vegetative
bioproductivity ~\cite{hardenberg_2001_prl}. To study these questions we introduce and analyze
a new model, which extends an earlier model for vegetated dunes~\cite{yizhaq_prl} to include crust dynamics.

The model consists of two coupled ordinary differential equations,
\begin{subequations}
\label{eq:model}
    \begin{eqnarray}
    \label{eq:vdot}
    \dot{v} & = & \alpha_v (v+\eta_v) s - \epsilon_v D_p
	v  g(v) s - \gamma D_p^{\frac{2}{3}}v \\ \nonumber
& & - \phi_v v b - \mu_v v \\
   \label{eq:bdot}
    \dot{b} & = & \alpha_b (b+\eta_b)
      s - \epsilon_b D_p b  g(v) s \\ \nonumber
& & - \phi_b b v - \mu_b b\ ,
   \end{eqnarray}
\end{subequations}

for the fractions $v$ and $b$ of surface cover by  vegetation and
biogenic crust, respectively,
where $s=1-v-b\ge 0$ represents the remaining bare sand, and the over
dot denotes the time derivative. The first terms on the right sides of
Eqs. (\ref{eq:vdot}) and (\ref{eq:bdot}) represent logistic
growth. Implicit in these growth forms is the assumption that the two
life forms, crust and vegetation, locally exclude one another; the presence of crust in a given location prevents the germination of plant seeds, while the presence of vegetation inhibits crust growth by blocking
sunlight. The growth rates of vegetation, $\alpha_v$, and of biogenic
crust, $\alpha_b$, are assumed to have the following dependence on
annual precipitation ($p$):
\begin{eqnarray}
   \label{eq:alpha}
   \alpha_i(p) = \left\{
     \begin{array}{lr}
       \alpha_\text{i,max}(1-e^{-(p-p_\text{i,min})/c_i}) &  p \geq
	p_\text{i,min} \\
       0 & p < p_\text{i,min} \,,
     \end{array}
   \right.
\end{eqnarray}
where $i=v,b$, $p_{\min}$ is a precipitation threshold below which there is no growth, and $\alpha_{i,\max}$ is the asymptotic growth rate at high precipitation levels. This form is in accordance with field observations \cite{danin_book}. The parameters $\eta_v$ and $\eta_b$ represent spontaneous growth rates due to, for example, a bank of seeds and spores in the soil, respectively.

Wind affects the vegetation and the crust both directly and indirectly. Indirect wind effects include sand transport, which leads to root exposure and burial of plants and crusts by sand. This process is represented by the second terms in (\ref{eq:model}a) and (\ref{eq:model}b). The parameter $D_p$, the \textit{drift potential}, is proportional to the potential bulk of sand that can be transported by the wind, and is given by \cite{bagnold}:
\begin{equation}
  \label{eq:D_p}
  D_p = \langle U^2(U-U_t) \rangle\,,
\end{equation}
where $U$ is the wind speed and $U_t$ is a threshold velocity that is necessary for sand transport (approximately 12 knots for wind measured at 10 m above the ground). If $U$ is measured in knots (1 knot $\approx$ 0.5 m/s), the units of $D_p$ are defined as \textit{vector units} ($VU$). $D_p$ can generally be classified into low, intermediate and high energy winds ($D_p < 200VU$, $200VU < D_p < 400VU$, and $D_p > 400VU$, respectively  \cite{fryberger}).

The function $g(v)$ introduces a wind shielding effect created by vegetation. Observations indicate \cite{wolfe_erosion_1993} that when vegetative cover exceeds a certain value ($v_c$), it induces a \textit{skimming flow} in which sand is protected from direct wind action. This value depends on various properties, such as plant shape and stem flexibility \cite{buckley_nature_1987,wiggs_kalahari_1995}. Based on these studies, we chose a continuous step-like function for $g(v)$:
\begin{equation}
 \label{eq:g}
 g(v) = \frac{1}{2} (\tanh (d(v_c-v)) + 1),
\end{equation}
such that $g \rightarrow 0$ for $v \gg v_c$ and $g \rightarrow 1$ for $v \ll v_c$. The sharpness of $g(v)$ is controlled by $d$. Since the indirect wind effect requires the availability of sand, the whole term is multiplied by $s$.

Direct wind effects are restricted, in the model, to vegetation and are represented by the third term in Eq. (\ref{eq:vdot}). This term accounts for stresses, such as increased evapotranspiration and branch cutting. It does not have a parallel in Eq. (\ref{eq:bdot}) since crust can sustain very intense winds \cite{belnap_vulnerability_1998}. Wind drag is proportional to the square of the wind velocity, and therefore, this term is proportional to $D_p^{2/3}$.

The parameters $\phi_v$ and $\phi_b$ represent non-local interactions between vegetation and crust not accounted for by the first terms of the RHS. It is still debated whether these interactions are positive or negative. On one hand, crust supports vegetation growth as a result of the ``source-sink'' effect \cite{zaady_runoff_1994}: water runoff from the crust (``source'') flowing toward vegetation patches (``sink''). On the other hand, crust suppresses vegetation by preventing water infiltration in the vicinity of the roots of the plants. Biogenic crust is usually suppressed by plants due to litter from nearby plants which limits light and may destroy the crust if the litter is toxic \cite{boeken_litter_2001}. Here, we assume that the negative relations are more significant. Finally, $\mu_v$ and $\mu_b$ represent grazing stress \cite{yizhaq_prl}.

The parameter values used in this study are based on Yizhaq et al. \cite{yizhaq_prl} for the equation of the vegetation dynamics (Eq. (\ref{eq:vdot})), and on the studies of Belnap et al. \cite{belnap_book,belnap_vulnerability_1998} for the crust (Eq. (\ref{eq:bdot})). The numerical values are:
  $\alpha_{v,max}=0.15$ $\left(\textrm{yr}^{-1}\right)$;
  $p_{v,min}=50 \left(\textrm{mm}\cdot\textrm{yr}^{-1}\right)$;
  $c_v=100 \left(\textrm{mm}\cdot\textrm{yr}^{-1}\right)$;
  $\eta_v=0.2$;
  $\epsilon_v=0.001 $ $ \left(\textrm{yr}^{-1} \cdot \textrm{VU}^{-1}\right)$;
  $\gamma = 0.0008 \left(\textrm{yr}^{-1} \cdot \textrm{VU}^{3/2}\right)$;
  $\phi_v=0.01 \left(\textrm{yr}^{-1}\right)$;
  $\mu_v=0 \left(\textrm{yr}^{-1}\right)$;
  $\alpha_{b,max}=0.015 \left(\textrm{yr}^{-1}\right)$;
  $p_{b,min}=20 \left(\textrm{mm}\cdot\textrm{yr}^{-1}\right)$;
  $c_b=50 \left(\textrm{mm}\cdot\textrm{yr}^{-1}\right)$;
  $\eta_b=0.1$;
  $\epsilon_b=0.0001 \left(\textrm{yr}^{-1} \cdot \textrm{VU}^{-1}\right)$;
  $\phi_b=0.01 \left(\textrm{yr}^{-1}\right)$;
  $\mu_b=0 \left(\textrm{yr}^{-1}\right)$;
  $v_c=0.3$;
  $d=15$.

\begin{figure}[ht]
  \centering
  \includegraphics[width=0.50\textwidth,height=6.0cm]{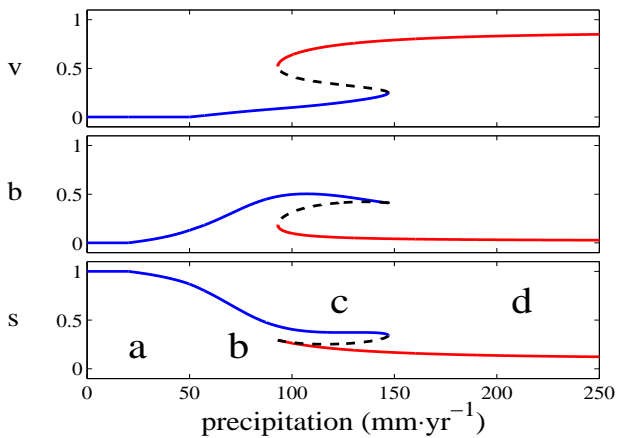}
  \caption{Bifurcation diagram of dune-cover states at low wind powers. Solid and dashed lines represent, respectively, stable and unstable steady solutions of Eqs. (\ref{eq:model}) with $D_p=120VU$, obtained using a numerical continuation method. The red and blue lines represent, respectively, the vegetation-dominated and the crust-dominated states. The labels a,b,c,d refer to the corresponding panels in Fig. \ref{fig:dunes_pgradient}. They represent bare dunes (a); mixtures of crust and vegetation forming semi-stable dunes (b) and stable dunes (c); and stable vegetated dunes (d).}
  \label{fig:p_bifu}
\end{figure}

The steady states of Eqs. (\ref{eq:vdot}) and (\ref{eq:bdot}) and
their stability properties for low wind powers are presented in the
bifurcation diagram shown in Fig. \ref{fig:p_bifu}. The results are
consistent with the general trend shown in
Fig. \ref{fig:dunes_pgradient} and reported in field
observations~\cite{belnap_book}: a low precipitation range (a) of bare
active dunes; intermediate precipitation ranges (b,c) of dunes with
mixed crust-vegetation coverage, semi-stabilized (b) or almost
stabilized (c); and a high precipitation range (d) of stabilized
vegetated dunes. In addition, the diagram predicts a bistability range
(c) of vegetation-dominated dunes ($v>b$) and crust-dominated
dunes ($b>v$). The bistability results from the negative vegetation-crust
interactions assumed in the model, which relies on the conjecture that a crusted soil prevents the germination
of plant seeds and also reduces the infiltration of surface water into
the soil, while a vegetated soil provides shading and possibly toxic
materials that inhibit the growth of crusts.

A different type of bistability is known to exist in regions of strong winds and high precipitation~\cite{tsoar_2005,yizhaq_prl}. This form of bistability results from the wind-shielding effect of the plants. The high wind power makes a bare dune active and prevents plant growth, despite the high precipitation level. However, once the dune is vegetated, the resulting lower wind power allows its persistence. Thus, two forms of bistability, designated here as Type I and Type II, are possible. Type I is associated with the wind-shielding effect of vegetation and occurs at high precipitation and strong winds. In this case, the alternative stable dune states are bare active dunes and vegetated fixed dunes.  Type II is associated with vegetation-crust competition and occurs at low precipitation and weak winds. Here, the alternative stable dune states are crust-dominated dunes and vegetation-dominated dunes. While bistability of Type I has been observed~\cite{tsoar_2005,yizhaq_prl}, observations of Type II have not yet been reported.

\begin{figure}[ht]
  \centering
  \includegraphics[width=0.45\textwidth,height=5.5cm]{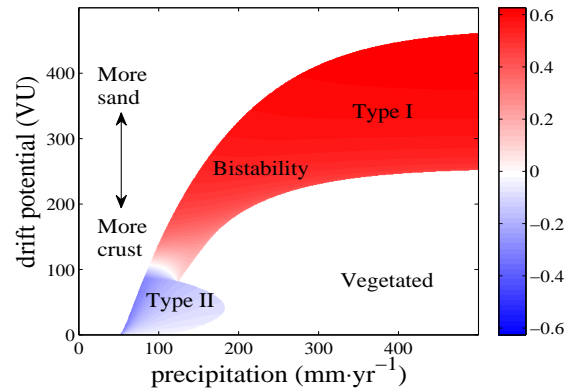}
  \caption{A phase diagram showing the monostability and bistability
    domains of dune-cover states. The shaded domain represents Type I
    bistability (active bare dune and fixed vegetated dune) at high
    precipitation and wind power values, and Type II bistability
    (fixed vegetation-dominated dune and fixed crust-dominated dune)
    at low precipitation and wind power values. The different shades
    represent the difference in sand-cover between the two stable
    states (calculated by subtracting the red line from the blue line
    in the bottom panel of Fig. \ref{fig:p_bifu}). The bistability
    domain is bounded by a monostability domain of vegetated dunes at
    high precipitation, and a monostability domain of unvegetated dunes
    at low precipitation, crusted at low wind power and bare at high wind power. }
  \label{fig:phase_diagram}
\end{figure}

Figure \ref{fig:phase_diagram} shows the domains of the two bistability forms in the plane spanned by the precipitation $p$ and the wind power $D_p$. The two domains are connected to form a continuous domain; proceeding from low to high $p$ and $D_p$ values, a cross-over from the bistability of Type II to Type I occurs. Bounding the continuous bistability domain are monostability domains of unvegetated dunes (bare or crusted) at low $p$ or high $D_p$, and vegetated dunes at high $p$ and low $D_p$.

\begin{figure}[ht]
  \centering
  \includegraphics[width=0.4\textwidth,height=5.5cm]{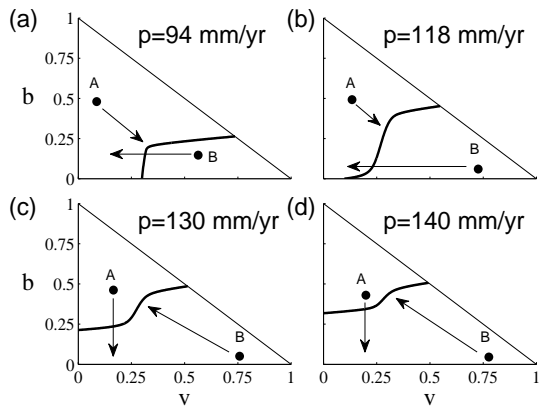}
  \caption{The phase space $v,b$ for the bistability of Type II at increasing precipitation values ($D_p=120VU$). The black dots A and B denote the crust-dominated state and the vegetation-dominated state, respectively. The line between the dots represents the boundary between the basins of attraction of the two alternative stable states. The arrows represent disturbances or manipulations capable of inducing state transitions (see main text for details).}
  \label{fig:phase_space}
\end{figure}

The existence of a biomass productive vegetation-dominated state and a less productive crust-dominated state, in the case of Type II bistability, implies the possible occurrence of desertification, i.e. a state transition inducing bioproductivity loss, as well as the feasibility of rehabilitation of vegetation, a state transition resulting in bioproductivity gain. By ``bioproductivity``, we refer to the total amount of vegetative biomass.
Such transitions can be triggered either by environmental variability,
for example, precipitation or wind-power fluctuations, or by anthropogenic
disturbances. The disturbance types that are necessary to trigger
desertification or the rehabilitation of vegetation can be determined by
examining the positions of the two stable states in relation to the
boundary between their basins of attraction, as
Fig. \ref{fig:phase_space} illustrates. Disturbances involving
vegetation removal can induce desertification (transition from point B
to point A) only at sufficiently low precipitation levels (panels (a) and
(b)). At higher precipitation levels (panels (c) and (d)), the disturbance
should also involve an increase in crust coverage (at the expense of
sandy soil), a rather unlikely disturbance scenario. Rehabilitation of
vegetation (transition from point A to point B) at relatively low
precipitation levels (panels (a) and (b)) cannot be triggered by crust
removal only -- planting is also necessary too. At higher precipitation levels (panels
(c) and (d)), crust removal alone can trigger such
rehabilitation. These conclusions may be affected by other choices for
the signs of the interaction coefficients $\phi_v$ and $\phi_b$.

The desertification form discussed above should be distinguished from that occurring in Type I bistability. In Type II bistability, both the productive and unproductive states (i.e. vegetated and crusted) represent stable, immobilized dunes, while in Type I, the non-vegetated (unproductive) state represents a mobile dune. Thus, desertification in the case of Type I bistability not only involves the loss of vegetation (bioproductivity) but may also lead to detrimental effects associated with dune mobility.

The results of the model qualitatively agree with a phenomenon that has been observed at the Israeli-Egyptian border, where sand dunes on the Egyptian side are active, while on the Israeli side, dunes are semi-stabilized. This difference in dune activity is the result of a vast cover of biogenic crust on the Israeli side and its absence on the Egyptian side. The phenomenon is clearly visible across the border line (see Fig. \ref{fig:dunes_pgradient}e) due to different albedo values for crust and sand. It was argued that biogenic crust is absent from the Egyptian dunes due to grazing activities that have led to crust trampling and erosion \cite{karnieli_1995}. Analyzing this region in our model ($p=100$ $\textrm{mm}\cdot\textrm{yr}^{-1}$), we find that grazing stress ($\mu_v=\mu_b=0.01$) yields bare dunes with low cover of crust and plants ($b=0.09, v=0.11 \rightarrow s=0.8$), while absence of grazing ($\mu_v=\mu_b=0$) yields crust-dominated dunes ($b=0.5, v=0.1 \rightarrow s=0.4$). Introducing grazing at low precipitation levels may, therefore, have a major effect on the fraction of bare dunes, doubling it in the numerical example presented above.

Further analysis of the model shows that steady mixed dune-cover states may undergo oscillatory instabilities. We find that if non-local competition is assumed to be negligible ($\phi_b=\phi_v=0$), the oscillations emerge in a supercritical Hopf bifurcation at $d=d_c$, where $d$ is the parameter that controls the wind-shield steepness in Eq. (\ref{eq:g})). As $d$ is increased beyond the Hopf bifurcation point, $d_c$, the amplitude of the oscillations increases and their frequency decreases. For $d \gg d_c$ (i.e. $g(v)$ approaches a step function), the amplitude saturates and the dynamics take the form of relaxation-oscillations.

We suggest the following mechanism for the oscillations. Let us begin
the description of a cycle with a state where $v>v_c$ and
$b<v_c$. Under such conditions, plants provide wind shield, which
promotes the growth of both vegetation and crust at the expense of
bare sand. As this growth proceeds, two processes become significant:
the diminishing bare sand, which slows down the growth of both
vegetation and crust; and the direct wind stress effect that
suppresses vegetation growth but does not affect the crust. The latter
process favors the growth of crust at the expense of vegetation. As a
result, the vegetation will reach a maximal value and start declining,
while the crust will continue growing. The growth of crust and the
decline of vegetation will continue until $v<v_c$. At this point, the
sand drift effect will become dominant, and the crust will decline
too, until there will be enough available sand for regrowth. Since the
growth rate of vegetation is significantly larger than that of crust,
vegetation will start growing at the expense of crust, until it exceeds $v_c$, and a new cycle will start over.

In summary, a mathematical model was introduced to analyze the effect
of biogenic crusts on sand dunes. Although simple, the model is able
to capture important aspects of the complex dynamics of biogenic
crusts and vegetation on sandy soils. Most significantly, it predicts
a new form of bistability in which the two alternative stable states
correspond to stabilized dunes with different proportions of
vegetation and crust coverage. This bistability form (Type II)
prevails at low precipitation and wind power values, and differs from the
bistability of bare dunes and vegetated dunes at high precipitation
and wind power values (Type I)~\cite{tsoar_2005,yizhaq_prl}. The two
bistability forms merge in the $p-D_p$ parameter plane to form a
single continuous domain with a small cross-over zone. The model sheds
new light on the vulnerability of sandy regions to desertification and
on the means to restore degraded vegetation.

We thank G. Bel, D. Perlstien, H. Tsoar, E. Zaady, and Y. Zarmi for fruitful discussions. We thank the Israel Science Foundation for financial support.


\end{document}